\documentstyle [12pt] {article}

\newcommand{\be}{\begin{equation}}
\newcommand{\ee}{\end{equation}}
\newcommand{\beqs}{\begin{eqnarray}}
\newcommand{\eeqs}{\end{eqnarray}}

\def\({\left(}
\def\){\right)}
\def\zlZ{{\zeta_2 \over \zeta_1}}
\def\zl{\zeta_1}
\def\zZ{\zeta_2}
\def\z{\zeta}
\def\d{\delta}
\def\D{\Delta}
\def\a{\alpha}
\def\b{\beta}

\def\na{\nabla}
\def\da{\dot{\alpha}}
\def\db{\dot{\beta}}
\def\pa{\partial}
\def\e{\epsilon}
\def\U{\Upsilon}
\def\S{\Sigma}
\def\th{\theta}

\def\ni{\noindent}

\begin{document}

\begin{titlepage}

\begin{flushright}
\begin{tabular}{l} ITP-SB-97-64 \\USITP-97-17\\ hep-th/9710250 \\ October, 1997
\end{tabular}
\end{flushright}

\vspace{8mm}
\begin{center} 

{\Large\bf Feynman rules in $N=2$ projective superspace I:}\\
\medskip
{\large\bf Massless hypermultiplets} 

\vspace{10mm}

F. Gonzalez-Rey \footnote{email: glezrey@insti.physics.sunysb.edu},
M. Ro\v{c}ek \footnote{email: rocek@insti.physics.sunysb.edu}, 
S. Wiles \footnote{email: swiles@insti.physics.sunysb.edu}\\
{\it Institute for Theoretical Physics,  State University of New York}\\
{\it Stony Brook, NY 11794, USA.\\}

\bigskip
U. Lindstr\"om \footnote{email: ul@vanosf.physto.se}\\
{\it Department of Physics, Stockholm University\\}
{\it Box 6730, S-11385 Stockholm, Sweden.\\}
{\it and\\}
{\it Institute of Physics, University of Oslo\\}
{\it Box 275, N-0314 Blindern, Oslo, Norway\\}

\bigskip
R. von Unge \footnote{email: unge@feynman.princeton.edu}\\
{\it Physics Department, Princeton University\\}
{\it Princeton, NJ 08544, USA.\\}

\vspace{10mm}

{\bf Abstract}
\end{center}

Manifestly $N=2$ supersymmetric Feynman rules are found for
different off-shell realizations of the massless 
hypermultiplet in projective superspace. When we reduce the Feynman rules
to an $N=1$ superspace we obtain the correct component
propagators. The Feynman rules are shown to be compatible with a
``duality'' that acts only on the auxiliary fields, as well as with
the usual duality relating the hypermultiplet to the tensor multiplet.

\vspace{25mm}

\end{titlepage}
\newpage
\setcounter{page}{1}
\pagestyle{plain}
\pagenumbering{arabic}
\renewcommand{\thefootnote}{\arabic{footnote}} \setcounter{footnote}{0}

\section{Introduction}

 Recently, there has been a great deal of interest in $N=2$
Super-Yang-Mills theory. A number of explicit computations have been
performed by reducing the $N=2$ tree level superspace action to $N=1$
component superfields whose Feynman rules are well known \cite{n1}. 
These calculations do not exhibit manifest $N=2$ supersymmetry and it 
is not possible to fully determine the form of the $N=2$ perturbative 
effective actions beyond the leading order terms. Here we present a
path integral quantization in $N=2$ superspace and construct very 
simple Feynman rules for the massless hypermultiplet in $N=2$ 
projective superspace\footnote{$N=2$ supersymmetric field 
theories with manifest $N=2$ Feyman rules and supergraphs exist  
in harmonic superspace \cite{gal}. Projective superspace is believed
to be related to the harmonic superspace, although the technology we
develop is quite different. The $N=2$ harmonic formalism has
been used recently to calculate interesting physical effects and low 
energy effective actions \cite{bu}.}. 

 There are many different off-shell realizations of the hypermultiplet which
all reduce to the same multiplet on-shell. All have either restricted
couplings or infinite numbers of auxiliary fields.  Here we focus on a class
of representations introduced in \cite{lr}, and compute the propagator and
vertices.  We first obtain the propagator in $N=1$ superspace and then
covariantize the result to find the corresponding $N=2$ superspace propagator 
; subsequently, we  derive the {\em same} result directly in $N=2$ superspace. 

 For simplicity, we restrict our analysis to the massless hypermultiplet, 
and leave the discussion of massive hypermultiplet \cite{massive}, 
couplings to the Yang-Mills multiplet \cite{n2_effec_action} and 
quantization of $N=2$ gauge multiplets \cite{gauge_quant} for the
future. We prove the known non-renormalization theorems
for the hypermultiplet.

\section{Projective Superspace}

 We begin with a brief review of projective superspace \cite{pss}.
The algebra of $N=2$ supercovariant derivatives in four dimensions 
is\footnote{ We will use the notation and normalization conventions of 
\cite{book}; in particular we denote $D^2 = {1 \over 2} D^\a D_\a$
and $\Box = {1 \over 2} \pa^{\a\da} \pa_{\a\da}$.}

\be
 \{ D_{a \a} , D_{b \b} \} = 0 \ , \ \  
 \{ D_{a \a} , \bar{D}^b_{\db} \} = i \d^b_a \pa_{\a\db } \  .
\label{n2_algebra}
\ee

\ni
We define an abelian subspace of $N=2$ superspace parameterized by 
a complex projective
coordinate $\z$ and spanned by the supercovariant derivatives

\beqs
 \na_\a (\z) & = & D_{1 \a} + \z D_{2 \a}\ , \\ 
 \bar{\na}_{\da} (\z) & = & \bar{D}^2_{\da} - \z \bar{D}^1_{\da}\ .
\eeqs

\ni
The conjugate of any object is constructed in this subspace by
composing the antipodal map on the Riemann sphere with
hermitian conjugation $\z^{\ast}\to-1/\z$ and multiplying by an
appropriate factor. For example,  

\be
 \bar{\na}_{\da} (\z) = \(- \z \) \( \na_\a \)^{\ast} \(-{1 \over \z}\)
 = \(- \z \) \( \bar{D}^{1}_{\da} + \(-{1 \over \z}\) \bar{D}^{2}_{\da}\)
\label{conjugation}
\ee

\ni
Throughout the paper, all conjugates of fields and operators in
projective superspace are defined in this sense. 

To make the global
$SU(2)$ transformation properties explicit we can introduce a
projective isospinor $u^a=(1,\z)$

\beqs
 \na_\a (\z) & = & u^a D_{a \a}\ , \\ 
 \bar{\na}_{\da} (\z) & = & \e_{ab} u^a \bar{D}^b_{\da}\ .
\eeqs

\ni 
Superfields living in this projective superspace obey the constraint

\be
 \na_{\a} \U = 0 = \bar{\na}_{\da} \U 
 \label{eta-constraint} \ ,
\ee

\ni
and the restricted measure of this subspace can be constructed from
the orthogonal operators $\D_\a=v^a D_{a \a},\bar{\D}_{\da} =\e_{ab}
v^a\bar{D}^b_{\da}$ where $v^a=(\z^{-1},-1), \e_{ab}u^a v^b = -2$. 
For constrained superfields, we
can write an $N=2$ supersymmetric action using this restricted measure 

\be 
 {1 \over 32 \pi i} \oint_C \z d \z d x \; \D^2 \bar{\D}^2 
 f(\U , \bar{\U}, \z)\ ,
\label{basic_action}
\ee

\ni
where $C$ is a contour in the $\z$-plane that generically
depends on $f$; in all the examples below, it will be a small contour
around the origin. Though our primary interest is in four dimensions, we
write the measure as $dx$ since the equations we write are valid for 
all $d \le 4$.

 The algebra that follows from (\ref{n2_algebra}) is 

\beqs
& \{ \na , \na \} = \{ \na , \bar{\na} \} = \{ \D , \D \} =
  \{ \D , \bar{\D} \} = \{ \na , \D \} = 0 &  , \nonumber \\
& &  \nonumber \\  & \{ \na_{\a} , \bar{\D}_{\da} \}
 = - \{ \bar{\na}_{\da} , \D_{\a} \} = 2i \pa_{\a \da}\ . &
\eeqs

\ni
For notational simplicity we write $D_{1 \a} =D_\a
, D_{2 \a} = Q_\a$. Then the identities

\be
 \D_\a = \z^{-1}(2 D_\a - \na_\a) \ , 
 \ \  \bar{\D}_{\da} = 2\bar{D}_{\da} + \z^{-1} \bar{\na}_{\da}\ , 
\ee

\ni
allow us to rewrite the action (\ref{basic_action}) in a form
convenient for reducing it to $N=1$ components:

\be 
 {1 \over 2 \pi i} \oint_C {d \z \over \z} \; d x \; D^2 \bar{D}^2 
 f(\U, \bar{\U}, \z)\ .
\label{eq-action}
\ee

The constraints (\ref{eta-constraint}) can be rewritten as 

\be
 D_\a \U = - \z Q_\a \U\  , \ \  \bar{Q}_{\da} \U = \z
 \bar{D}_{\da} \U\ .
\label{eta_constraint2}
\ee

\ni 
The superfields obeying such constraints may be classified \cite{lr}
as: i) $O(k)$ multiplets, ii) rational multiplets, iii) analytic
multiplets. We focus on $O(k)$ multiplets\footnote{$O(2)$ multiplets 
were first introduced in \cite{PLB147B} and the $O(k)$ generalization in 
\cite{ketov}. The harmonic superspace equivalent is given in 
\cite{ogiev}.}, which are
polynomials in $\z$ with powers ranging from $0$ to $k$, and on
analytic multiplets, which are analytic in some region of the Riemann
sphere. Later it will be useful to denote the $\z$ dependence of the
product $\z^i \times O(k)$ as $O(i,i+k)$.
 The transformation properties of the $O(k)$ multiplets under
global $SU(2)$ can be obtained from their parameterization in terms of
projective spinors and $SU(2)$ tensors

\be 
 \U = \sum_{n=0}^k \U_n \; \z^n \equiv u^{a_1} \dots u^{a_k} 
 L_{a_1 \dots a_k} ,\;\; \bar{\U} = \sum_{n=0}^{k} \bar{\U}_{n}
 \(-{1 \over \z} \)^{n}.  
\ee

\ni 
For even $k=2p$ we can impose a reality condition with respect the
conjugation defined above (see (\ref{conjugation})). We use $\eta$
to denote a real finite order superfield

\be 
 (-1)^p \z^{2p} \bar{\eta} = \eta \Longleftrightarrow 
 {\eta \over \z^p} = \overline{\( {\eta \over \z^p} \)} \ .
\label{eta_reality}
\ee

\ni 
This reality condition relates different coefficient superfields

\be 
 \eta_{2p-n} = (-)^{p-n} \bar{\eta}_n\ .  
\ee

\ni 
There are various types of analytic multiplets. The {\em arctic} multiplet
can be regarded as the limit $k \rightarrow \infty$ of the complex
$O(k)$ multiplet. It is analytic in $\z$, {\it i.e.}, around the north
pole of the Riemann sphere.

\be 
 \U = \sum_{n=0}^\infty \U_n \z^n \ .  
\ee

\ni 
Its conjugate superfield (the {\em antarctic} multiplet)

\be 
 \bar{\U} = \sum_{n=0}^\infty \bar{\U}_n (- {1 \over \z})^n \ , 
\ee

\ni 
is analytic in $\z^{-1}$, {\it i.e.}, around the south pole of the
Riemann sphere.

Similarly, the real {\em tropical}\footnote{We are happy to thank Warren
Siegel for suggesting this terminology.} multiplet is the limit $p
\rightarrow \infty$ of the real $O(-p,p)$ multiplet $\eta(2p)/
\z^p$. It is analytic away from the polar regions, and can be regarded
as a sum of a part analytic around the north pole and part analytic
around the south pole:

\be 
 V(\z) = \sum_{n=-\infty}^{+\infty} v_n \z^n \ , \ \ v_{-n}= (-)^n
 \bar{v}_n\ .  
\ee

The constraints (\ref{eta_constraint2}) relate the different
$\z$-coefficient superfields

\be 
 D_\a \U_{n+1}= - Q_\a \U_n \ , \ \ \bar{D}_{\da} \U_n =
 \bar{Q}_{\da} \U_{n+1}\ .  
\ee

\ni 
For any real $O(2p)$ multiplet these constraints are compatible
with the reality condition (\ref{eta_reality}). They also determine
what type of $N=1$ superfields the $\z$-coefficients are.

We illustrate this with the {\em real $O(4)$ multiplet}, which is the
first example we consider because it has the simplest Feynman
rules. Explicitly, it takes the form \cite{lr,pss}

\be 
 \eta = \bar{\Phi} + \z \bar{\S} + \z^2 X - \z^3 \S + \z^4 \Phi \ ,
 \ \ X = \bar{X}\ , 
\ee

\ni 
with $N=1$ superfield components:

\begin{itemize}
 \item[i)] an $N=1$ chiral superfield $\bar{D}_{\da} \Phi = 0$, also
           obeying $Q_\a \Phi = 0$; 
 \item[ii)] an $N=1$ complex linear superfield $\bar{D}^2 \S = 0$, 
            also obeying $Q^2 \S = 0$; 
 \item[iii)] an $N=1$ real unconstrained superfield $X$.
\end{itemize}

For the {\em complex $O(k)$ multiplet}, the coefficients corresponding
to the two lowest and highest orders in $\z$ are also $N=1$
constrained superfields, although they are not conjugate to each
other. All the intermediate coefficient superfields are unconstrained
in $N=1$ superspace

\be 
 \U = \bar{\Phi} + \z \bar{\S} + \z^2 X + \z^3 Y + \dots + \z^{p-1}
 \tilde{\S} + \z^p \tilde{\Phi} \ .
\ee

\ni 
Note that a complex $O(k)$ multiplet has twice as many physical
degrees of freedom as the real multiplet ({\it i.e.}, it describes two
hypermultiplets); nevertheless, we can write 

\be
 \eta = {1 \over  \sqrt{2} } (\U + (-\z^2)^p \bar\U ) 
\ee

\ni 
to construct a real $O(2p)$ multiplet out of a complex $O(2p)$ one. 
Conversely, we can write a complex $O(2p)$ multiplet out of {\em two} 
real multiplets $\eta,\tilde\eta$: 

\be
 \U = {\eta + i \tilde{\eta} \over \sqrt{2} }, \;\; 
 \bar{\U} = (-\z^2)^{-p} {\eta - i \tilde{\eta} \over \sqrt{2} } \ .
\ee

 For the arctic multiplet, only the two lowest coefficient
superfields are constrained. The other components are complex auxiliary
superfields unconstrained in $N=1$ superspace.

\be
 \U = \bar{\Phi} + \z \bar{\S} + \z^2 X + \z^3 Y + \dots 
\ee

Finally, for the real tropical multiplet all the $\z$-coefficient
superfields are unconstrained in $N=1$ superspace. 

\section{N=1 superspace description }

\subsection{N=1 actions}

 In the previous section, we defined a graded abelian subspace of
$N=2$ superspace and constructed
both a measure and constrained superfields that can be used to form $N=2$
invariant actions (\ref{eq-action}). 

For the real $O(2p)$ multiplet, the following action gives standard $N=1$
kinetic and interaction terms after performing the contour integral in $\z$
(although in general it is not $SU(2)$ invariant \cite{PLB147B}):

\be
 \int {d x \; D^2 \bar{D}^2} \oint {d\z \over 2 \pi i \z}\  
 \left[{1 \over 2} (-)^p \({\eta \over \z^p}\)^2 + 
 {\cal L}_I \({\eta \over \z^p}\) \right]\ .
\label{eq-Lagrangian}
\ee

\ni
Note that the natural variable for the function in the integrand of
(\ref{eq-Lagrangian}) is the self-conjugate superfield $\eta / \z^p$. 
Since the measure is already real, this field
allows us to construct manifestly real actions. The sign of the kinetic piece
guarantees that after performing the contour integration we obtain the right
kinetic terms for the chiral and linear component superfields. For the $O(4)$
multiplet, the free action in $N=1$ components is

\be
 \int {d x \; D^2 \bar{D}^2}\ ( \Phi \bar{\Phi} - \S \bar{\S} 
 + {1 \over 2} X^2 )\ .
\ee

\ni
We can use any real $O(2p)$ multiplet or the (ant)arctic multiplet to describe
the physical degrees of freedom of the $N=2$ hypermultiplet. The usual
description in terms of two $N=1$ chiral fields arises after a duality
transformation that replaces the $N=1$ complex linear superfield by a chiral
superfield. As usual, this can be done by rewriting the action in terms of a
parent action with a Lagrange multiplier $Z$, {\it e.g.}, for the $O(4)$ case,

\be
 \int {d x \; D^2 \bar{D}^2}\  ( \Phi \bar{\Phi} - S \bar{S} + {1 \over 2}
 X^2 + Z \bar{D}^2 \bar{S} + \bar{Z} D^2 S )\ . 
\label{n1_duality}
\ee

\ni
The field $S$ is unconstrained, but integrating out the field $Z$
imposes the linearity constraint on $S$. Alternatively we can integrate out
$S$ and recover the kinetic term of a chiral field $\bar{D}^2 Z$. The two
descriptions are dual formulations of the same physical degrees of
freedom. Except for the $O(2)$ multiplet, the duality transformation
merely changes the auxiliary fields of the theory, and for a nonlinear
$\sigma$-model, induces a coordinate transformation in target space.
The $O(2)$ case gives the four dimensional version of the well-known
$T$-duality.

For a complex $O(k)$ or (ant)arctic multiplet, a real action
necessarily involves both the field and its conjugate. The simplest
free action we can construct obeying hermiticity and $N=2$
supersymmetry is:

\be
 \int {dx D^2 \bar{D}^2} \oint {d\z \over 2 \pi i \z}\ \bar{\U} \U\ .
\label{eq-Lagrangian2}
\ee

\ni
The contour integration in the complex $\z$-plane produces the $N=1$
kinetic terms of the coefficient superfields for the complex $O(k)$ multiplet

\be
 \int {dx D^2 \bar{D}^2} \; \; \( \phi \bar{\phi} - \S \bar{\S}  + X
 \bar{X} - Y \bar{Y} + \dots + (-)^{k-1} \bar{\tilde{\S}} \tilde{\S} +
 (-)^k \bar{\tilde{\phi}} \tilde{\phi} \) \ ,
\ee

\ni
and for the (ant)arctic multiplet

\be
 \int {dx D^2 \bar{D}^2} \; \; \( \phi \bar{\phi} - \S \bar{\S}  + X
 \bar{X} - Y \bar{Y} + \dots \)\ .
\ee

\ni
Note that the kinetic terms for fields of even order in $\z$ have
opposite signs to those of odd order. Accordingly, the corresponding
$N=1$ propagators will also have opposite signs. Consequently, complex
$O(k)$ multiplets for odd $k$ contain chiral ghosts as highest
coefficient superfield, and are unphysical.  Nevertheless, a formal
calculation of an $N=2$ propagator is still possible, as we
see below. The large $k$ limit gives the (ant)arctic multiplet propagator
independently of whether $k$ is odd or even.

\subsection{N=1 Propagators}

To quantize the theory in $N=1$ components we need to know the
propagators of chiral, complex linear and auxiliary $N=1$ fields. The
last is trivial, while the first is well known. We briefly review the
calculation of the chiral field propagator to illustrate the general
techniques that we use to compute the propagator of the linear
superfield and the calculation in $N=2$ superspace.

 An $N=1$ chiral superfield $\Phi$ obeys the constraint $
\bar{D}_{\da} \Phi = 0 $. To calculate the propagator for a chiral
superfield we have the choice of adding either a constrained or an
unconstrained source term to the kinetic action \cite{book}. The
former involves a chiral integral 

\be
 \int dx \( \; \int d^4 \th \; \Phi \bar{\Phi} + \int d^2 \th \; j \Phi
 + \int d^2 \bar{\th} \; \bar{j} \bar{\Phi} \; \) \ .
\ee

\ni
We convert the chiral integrals into full superspace integrals and
rewrite the action with sources as

\be
 \int dx \int d^4 \th \( \Phi \bar{\Phi} + \Phi {D^2 \over \Box } j +
 \bar{\Phi} { {\bar{D}^2 \over \Box} \bar{j} } \) \ . 
\ee

\ni
Completing squares is now trivial

\be
 \int dx \int d^4 \th \left[ \( \Phi + { \bar{D}^2 \over \Box} 
 \bar{j} \) \; \( \bar{\Phi} + { D^2 \over \Box} j \) 
 - \bar{j} {1 \over \Box} j \right],
\label{complete_sq}
\ee

\ni
and the propagator is obtained by taking the functional derivative
with respect to the sources. The functional derivative with
respect to a chiral source $j$ is \cite{book}

\be 
 {\d j(x, \th) \over \d\ j'(x', \th')} = \bar{D}^2
 \d^4(\th-\th') \d (x-x')\ .
\ee

\ni
The antichiral-chiral propagator is therefore ${ - D^2 \bar{D}^2
\d / \Box}$.

We can also use the most general unconstrained source field $J$, with a
nonchiral coupling to the chiral field 

\be
 \int d^4 \th \left( J \Phi + \bar{J} \bar{\Phi}\right) = \int d^4 \th
 \left(  J {\bar{D}^2 D^2 \over \Box} \Phi + \bar{J} 
 {D^2 \bar{D}^2 \over \Box} \bar{\Phi}\right)\ ,
\ee 

\ni
where we have inserted projection operators for chiral and antichiral
fields. Since $j = \bar{D}^2 J$ is a solution to the chirality constraint of
$j$, we have effectively the same source coupling as above.
Completing squares we find

\be
 \int dx \: d^4 \th \left[ \( \Phi + { \bar{D}^2 D^2 \bar{J} \over \Box } \) 
 \( \bar{\Phi} + { D^2 \bar{D}^2 \bar{J} \over \Box } \) 
 - { \bar{J} D^2 \bar{D}^2 J \over \Box} \right]  \ ;
\ee

\ni
taking the functional derivative with respect to the 
source 

\be 
 {\d J(x, \th) \over \d\ J'(x', \th')} =
 \d^4(\th-\th') \d (x-x')\ ,
\ee

\ni
we obtain the same propagator

\be
 \langle \bar{\Phi} (1) \Phi (2) \rangle = -{ D^2 \bar{D}^2 \over \Box } \;
 \d^4( \th_1 - \th_2 )\d (x_1-x_2) \ . 
\ee

\ni
Constrained
sources have been used to derive Feynman rules only for chiral superfields.
For other fields we use unconstrained
sources and insert projectors corresponding to the subspace where the fields
live. This is how we compute propagators in $N=2$ superspace.

 Note that we include the supercovariant derivatives in the
propagator. Equivalently, we could put them into the interaction
vertices, and use $- \d^8_{12} / \Box$ as the propagator
\cite{book}.  A similar choice is possible for the constrained $N=2$
superfields we study.

 All this is well known $N=1$ technology. For the complex linear
superfield $\S$ the propagator has not appeared in the literature
(see, however, \cite{g}). We now derive this propagator using an
unconstrained superfield source $J$, and introduce suitable projectors
to complete squares in the action.

 The free action with sources is

\be
 \int d x \int d^4 \th \; (- \S \bar{\S} + \S \bar{J} +  J \bar{\S} )\ .
\ee

\ni
A linear superfield obeys the constraint $\bar{D}^2 \S = 0$, so we can
insert the projector $P = 1 - D^2 \bar{D}^2 / \Box$ and its conjugate in the
corresponding source terms. Integrating by parts and completing
squares we find 

\be
 \int d x \int d^4 \th \; \left[ -( \S - P J ) 
 ( \bar{\S}  - \bar{P} \bar{J} ) + J  \bar{P} \bar{J} \right] \ ,
\ee

\ni
which yields the propagator

\be
 \langle \bar{\S} (1) \S (2) \rangle\ =\ \bar{P} \; 
 \d^4 (\th_1 - \th_2 )\d (x_1-x_2) 
 = \(1 - { \bar{D}^2 D^2 \over \Box}\) \d_{12}\ .
\ee

\ni
An alternative way to compute the complex antilinear-linear propagator is to
perform the duality transformation in the $N=1$ component action {\em
with sources}: 

\be
 \int {d x \; D^2 \bar{D}^2} \; \; ( - S \bar{S} + Z \bar{D}^2 S  
 + \bar{Z} D^2 \bar{S} + J \bar{S} + S \bar{J} ) \ .
\ee

\ni
Completing squares and integrating out $S$ gives a dual action

\be
 \int {d x \; D^2 \bar{D}^2} \; \; ( \bar{D}^2 Z D^2 \bar{Z} + J \bar{J}  +
 \bar{J} D^2 \bar{Z} + \bar{D}^2 Z J )\ ;
\ee

\ni
we see that the complex antilinear-linear propagator is
equivalent to a chiral-antichiral propagator {\em plus} a contact term:

\beqs
 \langle \bar{\S} (1) \S (2) \rangle &=& {\d \over \d J_1} 
 {\d \over \d \bar{J}_2 } ln Z_0 \nonumber \\ \nonumber \\
 &=& \d^4 (\th_1 - \th_2 ) \d (x_1-x_2)  - 
 \langle \bar{D}^2 Z_1 D^2 \bar{Z}_2 \rangle \nonumber \\ \nonumber
 \\ &=& \( 1 - { \bar{D}^2 D^2 \over \Box} \) \d_{12}\ . 
\eeqs

\ni
Finally, the propagator for the real unconstrained superfield X is simply

\be
 -\d^4( \th_1 - \th_2) \d(x_1-x_2)  
\ee

\ni
because completing squares for an unconstrained superfield is
trivial.

\section{N=2 superspace propagators from $N=1$ component propagators} 

\subsection{The $O(4)$ multiplet}

For simplicity, we begin with the $O(4)$ multiplet.
We want to calculate the propagator $\langle \eta(\zl ,\th_1, p) \;
\eta(\zZ, \th_2, -p) \rangle$. Expanding in powers of $\zl$ and
$\zZ$, we can compute the two point functions of the $N=1$
coefficient superfields 

\beqs 
 \langle \eta (1) \; \eta (2) \rangle \mid _{\th_{\a}^2=0} & = & 
 \langle\bar{\Phi} (1) \Phi (2) \rangle \zZ^4+ \langle \Phi (1) 
 \bar{\Phi} (2) \rangle \zl^4 - \langle \bar{\S} (2) \S (2) \rangle 
 \zl \zZ^3           \nonumber \\ 
 \nonumber \\
& & - \langle \S (1) \bar{\S} (2) \rangle \zl^3 \zZ 
 + \langle X (1) X (2) \rangle \zl^2 \zZ^2  \ .
\label{eta propagator}
\eeqs

\ni
These are the only nonvanishing 2-point functions in the free
theory. Substituting the $N=1$ propagators from above, we find

\beqs 
 \langle \eta (1) \eta (2) \rangle|_{ \th_\a^2 = 0 } = & 
 \( -\zl\zZ(\zl^2+\zl\zZ+\zZ^2) - \zZ(-\zl^3 + \zZ^3) 
 { D^2 \bar{D}^2 \over \Box } - \zl( \zl^3 - \zZ^3 )
 {\bar{D}^2 D^2 \over \Box} \) & \nonumber \\
& \times  \d^4( \th_1 - \th_2 ) \d (x_1-x_2) \ . & 
\label{eta_prop2}
\eeqs

\ni
Given this form of the projected propagator, we try an ansatz for
the full $N=2$ propagator: by analogy to N=1 propagators for
chiral superfields, it should be proportional to $\na^4_1
\na^4_2$ acting on  $\d^8( \th_1 - \th_2 )$ ($\na_1^4 \equiv \na^2 (\zl)
\bar{\na}^2 (\zl) $). The $N=1$ projection of this is

\be
 \na^4_1 \na^4_2 \d^8( \th_1 - \th_2 ) \mid_{\th_{\a}^2=0} 
 = (\zl  - \zZ)^2 ( \zl^2 \bar{D}^2 D^2 + \zZ^2 D^2 \bar{D}^2 
 - \zl \zZ D \bar{D}^2 D ) \d^4( \th_1 - \th_2 )\ . 
\label{nabla-delta}
\ee

\ni
Rewriting (\ref{eta_prop2}) with the identity

\be 
 1 = {\bar{D}^2 D^2 \over \Box } + {D^2 \bar{D}^2 \over \Box } 
 - { D \bar{D}^2 D \over \Box } \ ,
\label{n=1_id}
\ee

\ni
and 

\be
 \zl^2 + \zl \zZ + \zZ^2 = {\zl^3 - \zZ^3 \over \zl - \zZ}\ ,
\ee

\ni
we find that the $N=1$ projection of 

\be 
 - { \zl^3 - \zZ^3 \over ( \zl - \zZ )^3 } \; \; 
 { \na_1^4 \na_2^4 \over \Box } \d^8( \th_1 - \th_2 ) \d (x_1-x_2) 
\label{n2_propeta}
\ee

\ni
reproduces (\ref{eta_prop2}). We note that the
coefficient of this $N=2$ propagator does not have a double pole at $\zl=
\zZ$ because

\be
 \na_1^4 \na_2^4 = \na_1^2 \( (\zl - \zZ )^2 \bar{D}^2 \bar{Q}^2 \) 
 \na_2^2\ .
\label{no_pole}
\ee

\ni
This ansatz for the $N=2$ propagator generalizes to the real
$O(2p)$ multiplet, and gives

\beqs 
 \langle \eta (1) \; \eta (2) \rangle \mid _{\th_{\a}^2=0} & = & 
 (-)^{p+1} \( \langle \bar{\Phi} (1) \Phi (2) \rangle \zZ^{2p} - 
 \langle \bar{\S} (1) \S (2) \rangle \zl \zZ^{2p-1} \right. \nonumber \\ 
\nonumber \\
& & \qquad \qquad + \langle \bar{X} (1) X (2) \rangle \zl^2 \zZ^{2p-2} 
 + \dots  \label{gral_real_prop} \\           
\nonumber \\
& & \qquad \qquad \left. - \langle \S (1) \bar{\S} (2) \rangle 
 \zl^{2p-1} \zZ + \langle \Phi (1) \bar{\Phi} (2) \rangle \zl^{2p} \) 
 \nonumber     \\ 
\nonumber \\ 
& = & (-)^{p+1} {\zl^{2p-1} - \zZ^{2p-1} \over ( \zl - \zZ )^3}\ 
 { \na_1^4 \na_2^4 \over \Box } \d^8( \th_1 - \th_2 ) \d (x_1-x_2) 
 \mid _{\th_{\a}^2=0}             \nonumber   \ . 
\eeqs

\ni
The real $O(2)$ multiplet has chiral, antichiral and real linear $N=1$
superfield coefficients. The propagator for the last can be obtained
by coupling to an unconstrained source and inserting the projector 
$-D \bar{D}^2 D / \Box$. 
The resulting real linear propagator is $- D \bar{D}^2 D \d / \Box$.
Despite the different off-shell degrees of freedom involved, the corresponding
$N=2$ propagator for the real $O(2)$ multiplet agrees with the general form
(\ref{gral_real_prop}) 

\beqs 
 \langle \eta (1) \; \eta (2) \rangle \mid _{\th_{\a}^2=0} & = & 
 { D^2 \bar{D}^2 \over \Box } \zZ^2 - { D \bar{D}^2 D \over \Box } 
 \zl \zZ + { {\bar{D}^2 D^2}\over \Box} \zl^2  \nonumber \\ 
\nonumber \\
& = & {1 \over \Box (\zl - \zZ)^2} \na_1^4 \na_2^4 \d_{12} 
 \mid _{\th_{\a}^2=0}\ . 
\label{o2}
\eeqs

\subsection{The complex $O(k)$ and (ant)arctic multiplets} 

Just as in the $O(4)$ case we can expand the propagator $\langle 
\bar{\U}_1 \U_2 \rangle$ in powers of $\zZ / \zl$ and compute the $N=1$
projection of the two point function 

\beqs
 \langle \bar{\U} (1) \; \U (2) \rangle \mid_{\th_{\a}^2=0} & = &
 \langle {\phi} (1) \bar{\phi} (2) \rangle - \(\zlZ\) \langle \S (1) 
 \bar{\S} (2) \rangle + \(\zlZ\)^2 \langle X (1) \bar{X} (2) \rangle 
  \nonumber 	\\ 
& & + \(-\zlZ\)^3 \langle Y(1)\bar{Y}(2) \rangle + \dots \nonumber \\ 
& & + \(-\zlZ\)^{k-1} \langle \bar{\tilde{\S}} (1) \tilde{\S} (2) \rangle 
  + \(-\zlZ\)^{k} \langle \bar{\tilde{\phi}}(1) \tilde{\phi}(2) \rangle \ .
\label{Upsilon_propagator}
\eeqs

\ni
As above, we insert the chiral-antichiral and linear-antilinear $N=1$
propagators in the two lowest order terms and the conjugate ones in the two
highest order terms of (\ref{Upsilon_propagator}). The $N=1$ auxiliary
superfields give contributions $\(\zZ /\zl\)^n\d_{12}(x) \d^4_{12}(\th)$.

For finite $O(k)$ complex multiplets we find

\beqs 
 \langle \bar{\U} (1) \; \U (2) \rangle \mid _{\th_{\a}^2=0} & = & 
 - {\bar{D}^2 D^2 \over \Box} \d_{12} - \zlZ 
 \(1 -{D^2 \bar{D}^2 \over \Box} \) \d_{12} - \(\zlZ\)^2
 \sum_{n=0}^{k-4} \(\zlZ\)^n \d_{12} \nonumber \\ 
& & - \(\zlZ\)^{k-1} \(1-{\bar{D}^2 D^2 \over \Box}\) 
 \d_{12} - \(\zlZ\)^k {D^2 \bar{D}^2 \over \Box} \d_{12} \nonumber \\ 
& = & \left[ \(1 - \(\zlZ\)^{k-1} \)\(-{\bar{D}^2 D^2 \over \Box} + 
 \zlZ {D^2 \bar{D}^2 \over \Box}\) - \sum_{n=1}^{k-1} 
 \(\zlZ\)^n \right] \d_{12} \nonumber	\\
& = & \left[ {\zl^{k-1} - \zZ^{k-1} \over \zl^k } \( \zl 
 {\bar{D}^2 D^2 \over \Box} + \zZ {D^2 \bar{D}^2 \over \Box} - 
 {\zZ \zl \over \zl - \zZ } \) \right] \d_{12}\ . 
\label{upsilo_prop}
\eeqs

For the (ant)arctic multiplet, we do not have 
$\bar{\tilde{\S}} \tilde{\S}$ and $\bar{\tilde{\phi}} \tilde{\phi}$,
so we have a geometric series in $\zZ / \zl$:

\beqs 
 \langle \bar{\U} (1) \; \U (2) \rangle \mid _{\th_{\a}^2=0} & = & 
 \left[ - {\bar{D}^2 D^2 \over \Box } - 
 \zlZ \(1 - {D^2\bar{D}^2 \over \Box }\) 
 - \sum_{n=2}^{\infty} \(\zlZ\)^n \right] \d_{12} \nonumber \\ 
& = & \left[ - {\bar{D}^2 D^2 \over \Box } + \zlZ 
 {D^2 \bar{D}^2 \over \Box} - \sum_{n=1}^{\infty} \(\zlZ\)^n \right] 
 \d_{12}\ .
\eeqs

\ni
In the region where $|\zZ / \zl| < 1$ the series is convergent to 
$\zZ / (\zl -\zZ)$. If we use this propagator to connect
$\bar{\U}_1 \U_2$ lines from vertices at different points in the
$\z$-plane and form a closed loop, when performing the contour integrals
of each vertex the pole at $\zl=\zZ$ can lead to ambiguities\footnote{If the
vertices of the theory depend on a real combination $\z^p\bar{\U} +(-\z)^{-p}
\U$ for any $p$, the potential problem with the summed series disappears. We
will see that such a dependence allows us to perform a duality transformation
between the real $O(2p)$ multiplet action and the analytic multiplet
action.}. 

Just as we did in the real $O(4)$ multiplet case, we find the $N=2$ 
propagator of
the complex $O(k)$ multiplet by using an ansatz to guess the $N=2$
expression whose $N=1$ reduction reproduces our result:

\beqs 
 \langle \bar{\U} (1) \; \U (2) \rangle & = & - {\sum_{n=0}^{k-2}
 \zZ^n \zl^{k-2-n} \over \zl^k ( \zl - \zZ )^2 } 
 \  {\na_1^4 \na_2^4 \over \Box } \d^8 (\th_1 - \th_2)
 \d (x_1-x_2) \label{n2_proplin12}	\\ 
& = & - { \zl^{k-1} -
 \zZ^{k-1} \over \zl^k ( \zl - \zZ )^3 } 
 \; \; { \na_1^4 \na_2^4 \over \Box } \d^8 (\th_1 - \th_2)
 \d (x_1-x_2) \label{n2_proplin12'}
\eeqs

\ni
and its conjugate

\be 
 \langle \U (1) \; \bar{\U} (2) \rangle = - {\zl^{k-1} - \zZ^{k-1} 
 \over \zZ^k ( \zl - \zZ )^3 } \; \; 
 { \na_1^4 \na_2^4 \over \Box } \d^8( \th_1 - \th_2 )
 \d (x_1-x_2) \ .
\label{n2_proplin21}
\ee

\ni
This result is consistent with our previous observation that a real $O(2p)$
superfield can be constructed from a complex $O(2p)$ multiplet and its
conjugate:

\beqs 
 \langle \eta (1) \eta (2) \rangle & = & {1 \over 2} \langle 
 \( \U (1) + (-)^p \zl^{2p} \bar{\U} (1) \) \; 
 \( \U (2) + (-)^p \zZ^{2p} \bar{\U}(2) \) \rangle  \nonumber	\\ 
& & \nonumber \\
& = &{1 \over 2} \left[ (-)^p \zZ^{2p} \langle \U (1) \bar{\U} (2) \rangle + 
 (-)^p \zl^{2p} \langle \bar{\U} (1) \U (2) \rangle \right] \ ; 
\eeqs

\ni
likewise, the propagator of the complex $O(2p)$ multiplet and
its conjugate can be reobtained from that of a real $O(2p)$ superfield by
complexification.

The (ant)arctic multiplet propagator can also be reconstructed in this
form and we find

\beqs 
 \langle \bar{\U} (1) \; \U (2) \rangle & = & - {1 \over \zl^2} 
 \sum_{n=0}^{\infty} \(\zlZ\)^n \ {\na_1^4 \na_2^4
 \over (\zl -\zZ)^2 \Box} \d^8( \th_1 - \th_2 ) \d (x_1-x_2) \nonumber\\
& = & - {1\over\zl(\zl-\zZ)^3} \ {\na_1^4\na_2^4\over\Box} 
 \d^8(\th_1-\th_2) \d (x_1-x_2)\ , \label{n2_propupsi12} \\ 
& & \nonumber \\
 \langle \U (1) \; \bar{\U} (2) \rangle & = & - { 1 \over \zZ^2 } 
 \sum_{n=0}^{\infty} \( {\zl \over \zZ} \)^n \; \; { \na_1^4 \na_2^4
 \over (\zl-\zZ)^2 \Box } \d^8(\th_1 -\th_2)\d (x_1-x_2) \nonumber\\
& = & { 1 \over \zZ ( \zl - \zZ)^3 } \; \; { \na_1^4 \na_2^4 \over \Box }
 \d^8( \th_1 - \th_2 ) \d (x_1-x_2)\ .
\label{n2_propupsi21}
\eeqs

\section{Feynman rules derived in $N=2$ superspace}

\subsection{$N=2$ Propagators} 

We now derive the $N=2$ propagator of the complex $O(k)$ multiplet 
directly in $N=2$ superspace and validate our ansatz. The real $O(2p)$
multiplet propagator can be obtained using very similar arguments and
does not provide any additional information. We introduce the operators

\beqs
  M(\z) &=&  {1 \over 16\Box^{2}} \left(
             \D^{4} + {1 \over \z}\na^{\a}\D_{\a}\bar{\D}^2
             +{1 \over \z^2}\na^{2}\bar{\D}^{2} \right)
         =   {1 \over 4\Box^{2}} {D^2 \bar{\D}^2 \over \z^{2}}
                , \nonumber \\
  N(\z) &=&  {1 \over 16\Box^{2}} \left(
             \D^{4} - {1 \over \z}\na^{\a}\D_{\a}\bar{\D}^2
             +{1 \over \z^2}\na^{2}\bar{\D}^{2} \right)
         =   {1 \over 4\Box^{2}} Q^{2} \bar{\D}^{2}.
\eeqs

\ni
that satisfy the following relations

\beqs
 \na^4_1 M(\zZ) \na^4_2 & = & { \( {\zl \over \zZ } \)^2
   \over (\zl - \zZ)^2 } {\na^4_1 \na^4_2 \over \Box}       \nonumber\\
 \na^4_1 N(\zZ) \na^4_2 & = & {1 \over \( \zl - \zZ \)^2 } 
 {\na^4_1 \na^4_2 \over \Box} \ .
\label{nice_relations}
\eeqs

\ni
Inspired by the analogy to $N=1$ chiral superfields,
we introduce an unconstrained prepotential superfield and its
conjugate

\be
 \U = \na^4 \psi = \na^4 \sum_{n=0}^{k-4} \psi_n \z^n, \;\; 
 \bar{\U} = {\na^4 \over \z^4} \bar{\psi} = {\na^4 \over \z^4}
 \sum_{m=0}^{k-4} \bar{\psi}_{-m} \z^{-m} \ .
\ee

\ni
In terms of $\psi$, the
kinetic action can be rewritten with the full $N=2$ measure 

\be
 \int dx D^2 \bar{D}^2 \oint {d \z \over 2 \pi i \z} \; \; 
 \na^4 \( {\na^4 \over \z^4} \bar{\psi} \; \psi \) = \int dx \;
 d^8\th \oint {d \z \over 2 \pi i \z} \;\; {\na^4 \over \z^2} 
 \bar{\psi} \; \psi \ .\label{prepot_kin}
\ee

\ni
We derive the complex $O(k)$ and (ant)arctic
multiplet propagators by completing squares in the full $N=2$ action with
sources. The simplest source is an unconstrained $O(k)$
superfield and its conjugate, which gives the action

\be
 S_0+S_J=\int dx \; d^8 \th \oint {d \z \over 2 \pi i \z } \(\bar{\psi} 
 {\na^4 \over \z^2} \psi + \bar{J} \U + \bar{\U} J \) \ .
\label{analy-source}
\ee

\ni
When we reduce this action to $N=1$ components, every coefficient
superfield $\U_n$ couples to an unconstrained source. The functional
derivative of $\z$-dependent $N=2$ superfields is defined by

\be 
 {\d \over \d J(x',\th',\z')} \int dx \: d^8 \th 
 \oint {d \z \over 2 \pi i \z} J F = F(x',\th',\z')\ .
\label{riemann_delta_def}
\ee

\ni
For a generic $O(i,j)$ source $J=\sum_{n=i}^j\z^n J_n$,

\be 
{\d \over \d J(x',\th',\z')} J(x,\th,\z)  =  \d_{(i)}^{(j)}(\z,\z') 
 \d^{8}(\th - \th^{\prime})\d(x-x') \ ,
\label{gral_f_der}
\ee

\ni
where the delta function on the Riemann sphere 

\be
 \d_{(i)}^{(j)}(\z,\z') = \sum_{n=i}^{j} \({\z \over \z'}\)^n  
 = {\z^i \over (\z')^j} {(\z')^{j-i+1} - \z^{j-i+1} \over \z' - \z }
 = \d_{(-j)}^{(-i)}(\z',\z)
\ee

\ni
projects onto the subspace of $O(i,j)$ functions when integrated with
the measure $\oint{ d\z' \over 2\pi i\z'}$. Note that 
$\d_{(i)}^{(j)}(\z,\z')$ is a function that approaches a well defined
distribution when $i$ and (or) $j$ tends to infinity.
For the $O(k)$ source in our action, the functional derivative
 
\be
 {\d \over \d J(x',\th',\z')} J(x,\th,\z)  = \d_{(0)}^{(k)}(\z,\z')
 \d^{8}(\th - \th^{\prime})\d(x-x')  \ .
\label{Ok_f_der}
\ee

\ni
acting on the free theory path integral will give us the propagators.
We now rewrite the action with sources (\ref{analy-source}) in terms
of the prepotential

\be 
S_0 + S_J = \int dx \; d^8 \th \oint {d \z \over 2 \pi i\z } 
 \left(\bar{\psi} {\na^4 \over \z^2} \psi + \bar{J} \na^4 \psi + \bar{\psi}
 {\na^4 \over \z^4} J\right)\ .
\label{preppie}
\ee

\ni
We cannot complete squares directly on the prepotential\footnote{Although 
the prepotential has a gauge invariance whose fixing may introduce
ghosts for ghosts, we will not be concerned with it here, as in the
absence of nonabelian gauge fields the
ghosts decouple. See, however, \cite{g}.} because even the
simplest sources have a different $\z$-dependence than the
prepotentials.  Hence, we must insert a projector with the
following properties : i) it leaves the source coupling to
$\bar{\psi}$ ($\psi$) invariant; ii) it can be split in the
product of the kinetic operator and some inverse operator that
projects the source into the subspace of $O(k-4)$ functions.

We will proceed as follows: The identity $16\Box^{2}\na^{4} =
\na^{4}\D^{4}\na^{4}$ defines a projection operator in $N=2$
projective superspace. We may use it to rewrite the last source term as

\be
 {\cal L}_{J} = \oint {d\z\over 2 \pi i \z} \bar{\psi}{\na^{4} \over
    \z^{4}}J = 
 \oint {d\z\over 2 \pi i \z} \bar{\psi}{\na^{4}\D^{4}\na^{4}
    \over 16\Box^{2}\z^{4}}J \ .
\ee

\ni
However, since the operators $\na^2 (\z) \bar{\D}^2 (\z)$ and 
$\na^\a (\z) \D_\a (\z) \bar{\D}^2 (\z)$ are annihilated when inserted
between two $\na^4 (\z)$ operators,  we may add any combination
of those to the $\D^{4}$ operator. In particular we rewrite the source
term as

\be
 {\cal L}_J = \oint {d\z\over 2 \pi i \z} \bar{\psi}{\na^{4} \over \z^{2}}
   M(\z) {\na^{4} \over \z^{2}}J  =
 \oint {d\z\over 2 \pi i \z} \bar{\psi}{\na^{4} \over \z^{2}}
   N(\z) {\na^{4} \over \z^{2}}J \ .
\ee

\ni
Because $\bar{\psi}{\na^{4} \over \z^{2}}$ is $O(-k+2,2)$, we can
insert a $\d_{(-2)}^{(k-2)}$ and split the source term:

\be
 {\cal L}_J= \oint {d\z \over 2 \pi i \z } \oint {d\z' \over 2
    \pi i \z'} \bar{\psi} (\z) {\na^4 (\z) \over \z^2 } \;
   \d_{(-2)}^{(k-2)}(\z,\z') M(\z') {\na^4 (\z') \over \z'^2} J (\z') \ .
\ee

\ni
To complete
squares we still have to prove that the source is projected into a
$O(0,k-4)$ field. Using the relations (\ref{nice_relations}) we may rewrite
the source term once more

\be
 {\cal L}_J = \oint {d \z \over 2 \pi i \z} \oint {d \z' \over 2 \pi
     i \z' } \bar{\psi} (\z) {\na^4 (\z) \over \z^2 } \;
  \d_{(0)}^{(k)}(\z,\z') N(\z') {\na^4 (\z') \over \z'^2} J (\z') \ ,
\ee

\ni
where the extra factor of $(\z / \z')^2$ in
(\ref{nice_relations}) effectively shifts the range of the delta
function two steps so that it now runs from $0$ to $k$. However,
$\bar{\psi}{\na^4 \over \z^2}$ is still $O(-k+2,2)$ so the remaining
contribution is

\be
 {\cal L}_J = \oint {d\z \over 2 \pi i \z} \oint {d\z' \over 2 
   \pi i \z'} \bar{\psi} (\z) {\na^4 (\z) \over \z^2 } \;
  \d_{(0)}^{(k-2)} (\z,\z') N(\z') {\na^4 (\z') \over \z'^2 } J (\z') \ . 
\ee

\ni
To see that there are no higher powers of $\z$ than $k-4$, we write

\be
 {\cal L}_J = \oint {d \z \over 2 \pi i \z } \oint {d \z' \over 2 \pi
 i \z' } \bar{\psi} (\z) {\na^4 (\z) \over \z^2 } \left[
 \d_{(0)}^{(k-4)} (\z,\z') N(\z') + \d_{(k-3)}^{(k-2)}(\z,\z') N(\z') \right]
 {\na^4 (\z') \over \z'^2} J (\z'),
\ee

\ni
and use the relations (\ref{nice_relations}) on the last term to get

\be
 {\cal L}_J = \oint {d \z \over 2 \pi i \z} \oint {d \z' \over 2 \pi
 i \z'} \bar{\psi} (\z) {\na^4 (\z) \over \z^2} \left[
 \d_{(0)}^{(k-4)}(\z,\z') N(\z') + \d_{(k-5)}^{(k-4)}(\z,\z') M(\z') \right]
 {\na^4 (\z') \over \z'^2} J (\z'),
\ee

\ni
which proves that $\bar{\psi} {\na^4 \over \z^2}$ couples to a
$O(k-4)$ projected source:

\be
 {\cal J}^{(k-4)}(\z) \equiv \oint {d \z' \over 2 \pi i \z'} \left[
  \d_{(0)}^{(k-4)}(\z,\z') N(\z') + \d_{(k-5)}^{(k-4)}(\z,\z') M(\z') \right]
 {\na^{4} (\z') \over \z'^2 } J (\z') \ .
\ee

\ni
Note that

\be
 \na^{4} {\cal J}^{(k-4)}(\z) =
  \na^{4} \oint {d \z' \over 2 \pi i \z' } \;
  \d_{(0)}^{(k-2)}(\z,\z') N(\z') {\na^4 (\z') \over \z'^2 } J (\z') \ .
\ee

\ni
Similarly

\be
 \oint {d \z \over 2 \pi i \z} \psi \na^4 \bar{J} = 
 \oint {d \z \over 2 \pi i \z} \psi {\na^4 \over \z^2} \bar{\cal J}^{(4-k)}
\ee

\ni
where 

\be
 \bar{\cal J}^{(4-k)} (\z) = \oint {d \z' \over 2 \pi i \z'} \;\; 
 \left[ \d_{(4-k)}^{(0)} (\z, \z') M(\z') + \d_{(4-k)}^{(5-k)} (\z, \z') 
 N(\z') \right] \z'^2 \na^4 (\z') \bar{J} (\z')
\ee

\ni
with

\be
 \na^{4} \bar{\cal J}^{(4-k)}(\z) = 
  \na^{4} \oint {d \z' \over 2 \pi i \z'} \;\; 
  \d_{(2-k)}^{(0)} (\z, \z') M(\z') \; \z'^2 \na^4 (\z') \bar{J} (z') \ .
\ee

Since we have shown that the prepotential and the projected source
term are of the same type, we may now complete squares in the action
(\ref{preppie})

\beqs
 S_{0} + S_{J} = \int dx d^{8}\th \oint {d \z_0 \over 2 \pi i \z_0} 
 \left[ \left(\bar{\psi} (\z_0) + \bar{{\cal J}}^{(4-k)} (\z_0) \right)
     {\na^{4} (\z_0) \over \z_0^2}
  \left(\psi (\z_0) + {\cal J}^{(k-4)} (\z_0) \right) \right. & &
\eeqs

\beqs
- \left. \oint \oint {d \z d \z'\over (2 \pi i)^2 \z \z'}  
 \left( \d_{(2-k)}^{(0)} (\z_0,\z) M(\z) \z^2 \na^4 (\z) \bar{J} (\z) \right) 
 {\na^4 (\z_0) \over \z_0^2}
 \left( \d_{(0)}^{(k-2)} (\z_0,\z') N(\z') {\na^4 (\z') \over \z'^2} 
 J (\z') \right) \right]. & & \nonumber
\eeqs

\ni
All the source dependence in the first term may be absorbed in a
redefinition of the prepotential since they are both $O(k-4)$
multiplets. Using the same arguments backwards, we can rewrite 
the term quadratic in sources

\beqs
ln Z_{0}[J,\bar{J}] & = & - \int dx d^8 \th \oint {d\z \over 2 \pi i \z} 
 \oint {d \z_0 \over 2 \pi i \z_0} \left( \d_{(2-k)}^{(0)} (\z,\z_0) 
 \bar{J} (\z_0) \na^4 (\z_0) M(\z_0) \z_0^2 \right) {\na^4 (\z) \over \z^2}
 {\cal J}^{(k-4)} (\z)  \nonumber    \\
& = & - \int dx d^8 \th \oint {d\z \over 2 \pi i \z} \bar{J} (\z)
 {\na^{4} (\z) \D^{4} (\z) \na^{4} (\z) \over 16\Box^{2} } 
 {\cal J}^{(k-4)} (\z)  \\
& = &  -\int dx d^8 \th \oint {d\z \over 2 \pi i \z} 
 \oint {d \z' \over 2 \pi i \z'} 
 \bar{J} (\z) \na^4 (\z) \; \d_{(0)}^{(k-2)} (\z,\z') N(\z')
   {\na^4 (\z') \over \z'^2} J (\z') \nonumber  .
\eeqs

\ni
Using (\ref{nice_relations}) once more, we finally arrive at:

\be
 Z_{0}[J,\bar{J}] = \exp \left( -\int dx d^8 \th 
  \oint {d\z \over 2 \pi i \z} \oint {d\z' \over 2 \pi i \z'} 
 \bar{J} (\z) \; \d_{(0)}^{(k-2)}(\z,\z')
   {\na^4 (\z) \na^4 (\z') \over \Box \z'^2 (\z -\z' )^2 } J (\z')
 \right).
\label{ok_path_int}
\ee

\ni
The complex $O(k)$ multiplet propagator is obtained by
functionally differentiating with respect to the unconstrained $N=2$ sources

\beqs
\langle \bar{\U} (1) \U (2) \rangle & = & 
   {\d^2 \over \d J (\z_1) \, \d \bar{J} (\z_2)} \ln Z_0
 \nonumber \\ \nonumber \\
& = &  \oint {d\z \over 2 \pi i \z}
  \oint {d\z' \over 2 \pi i \z'} \;
  \d_{(0)}^{(k)}(\z_1,\z') \;
  \d_{(-k)}^{(0)}(\z_2,\z) \;
  \d_{(0)}^{(k-2)}(\z,\z')
\nonumber\\
&{ }&\qquad\qquad\qquad\qquad \times\quad  
 {\na^{4} (\z) \na^{4} (\z') \over \z'^{2}(\z-\z')^2 \Box} \;
  \d^{8}(\th - \th') \; \d(x-x').
\eeqs

\ni
Evaluating the $\z$ and $\z'$ integrals after using (\ref{no_pole}) to
cancel the double pole in $\z - \z'$, we obtain our previous result
(\ref{n2_proplin12}). Note that this can be rewritten as

\be
 \langle \bar{\U}(1) \U(2) \rangle =
 - \d_{(0)}^{(k-2)}(\zZ,\zl) {\na_1^4 \na_2^4 \over \zl^2 (\zl-\zZ)^2 \Box}
 \;  \d^{8}(\th_1 - \th_2) \; \d(x_1 - x_2) \ .
\ee 

The limit $k \rightarrow \infty$ gives the
(ant)arctic multiplet propagator. The derivation in
$N=2$ superspace follows exactly the same lines as the derivation for
the complex $O(k)$ multiplet. The only difference is that the
prepotential $\bar{\psi}$ is now antarctic. This implies that 
only the two lowest powers of $\z$ in the projected source 
present a problem, the higher power being infinite. 
Therefore rewriting the projected source in terms of the
operator $N(\z_2)$ is enough to obtain an arctic function coupled to 
$\bar{\psi}(\z_1) \na_1^4 / \z_1^2$.
The propagator for the arctic multiplet derived in this
way agrees with the form of the propagator found earlier in
(\ref{n2_propupsi12}) and (\ref{n2_propupsi21}).

\bigskip
For completeness we will also calculate the propagator of the real
$O(2)$ multiplet in $N=2$ superspace. A solution to the constraint
(\ref{eta-constraint}) obeying the reality condition and with the correct
global $SU(2)$ transformation properties is:

\be
\eta = \na^4 ( \D^2 + \bar{\D}^2 ) \Psi \ .
\ee

\ni
The product $\na^2 \D^2 = 4 D^2 Q^2$ is $\z$-independent, and
therefore $\Psi$ is a $\z$-independent dimensionless isoscalar. We can
rewrite the free action with sources in terms of this prepotential 

\be 
S_0 + S_J = \int dx \; d^8 \th \oint {d \z \over 2 \pi i \z} 
 \( {1 \over 2} \Psi ( \D^2 + \bar{\D}^2 ) \na^4 (\D^2 + \bar{\D}^2 ) \Psi
 + \Psi (\D^2 + \bar{\D}^2 ) \na^4 {J \over \z^2} \)\ ,
\label{eta2_source}
\ee

\ni
and complete squares in $\Psi$. This is particularly easy in this
case because the kinetic operator of the prepotential acts as $\Box^2$
on the prepotential source

\be
 \( \D^2 + \bar{\D}^2 \) \na^4 \( \D^2 + \bar{\D}^2 \)  
 (\D^2 + \bar{\D}^2 ) \na^4 J = 32 \Box^2 (\D^2 + \bar{\D}^2 ) 
 \na^4 J   \ ,
\ee

\ni
and in addition this kinetic operator is $\z$-independent

\be
 \( \D^2 + \bar{\D}^2 \) \na^4 \( \D^2 + \bar{\D}^2 \) = 
 16 D^2 Q^2 \bar{D}^2 \bar{Q}^2 + 16 \bar{D}^2 \bar{Q}^2 D^2 Q^2 +
 16 D^2 Q^2 \Box + 16 \bar{D}^2 \bar{Q}^2 \Box \ .
\ee

\ni
Defining the following $\z$-independent source

\be
{\cal J} = \oint {d \z \over 2 \pi i \z} (\D^2 + \bar{\D}^2 ) 
 \na^4 {J(\z) \over 32 \Box^2 \z^2}
\ee

\ni
the path integral can be expressed as a completed square

\beqs
ln Z_0[J] & = & \int dx \; d^8 \th {1 \over 2} ( \psi + {\cal J} )    
 ( \D^2 + \bar{\D}^2 ) \na^4 (\D^2 + \bar{\D}^2 ) 
 ( \psi + {\cal J} )                           \nonumber   \\
\nonumber  \\ 
& & - {1 \over 2} \oint {d \z_1 \over 2 \pi i \z_1} 
  \oint {d \z_2 \over 2 \pi i \z_2} {J(\z_1) \over \z_1^2} \na_1^4 
 (\D_1^2 + \bar{\D}_1^2 ) {1 \over 32 \Box^2} (\D_2^2 + \bar{\D}_2^2 ) 
 \na_2^4 {J \over \z_2^2} .  
\eeqs

\ni
The resulting propagator is

\be 
 \langle \eta (1) \eta (2) \rangle = \na_1^4 (\D_1^2 + \bar{\D}_1^2) 
 ( \D_2^2 + \bar{\D}_2^2 ) \na_2^4 { \d_{12} \over 32 \Box^2}  
 = {1 \over \Box (\zl - \zZ)^2} \na_1^4 \na_2^4 \d_{12}\ ,
\ee

\ni
in agreement with our previous calculation (\ref{o2}).

\subsection{Vertices}

We now define vertex factors that allow us to construct
the diagrams of the interacting theory. As mentioned before, we can
put the graded spinor derivatives $\na^4_1 , \na^4_2$ of a propagator
$\langle \eta (1) \eta (2) \rangle$ on the internal lines of interaction 
vertices. For the real $O(2p)$ multiplet, we would be working  formally 
with propagators

\be
 \langle \eta (1) \eta (2) \rangle = (-)^{p+1} 
 { \zl^{2p-1} - \zZ^{2p-1} \over ( \zl - \zZ )^3} \ 
 { 1 \over \Box } \d^8( \th_1 - \th_2 ) \d (x_1-x_2)
\ee

\ni
connecting two $\eta$ internal lines. Accordingly, interactions 
such as

\be
\int dx \; d^4 \th \oint {d \z \over 2 \pi i \z} \; {1 \over n!} 
 \( {\eta \over \z^p} \)^n
\ee

\ni
in which $q$ lines are external and $n-q$ lines are internal will 
contribute (with appropriate combinatorial normalization) as vertex 
factors of the form 

\be
\int d^4 \th \oint {d \z \over 2 \pi i \z} \( {\eta \over \z^p} \)^q 
 \( {\na^4 \over \z^p }\)^{n-q} \ .
\ee

\ni 
For the complex $O(k)$ multiplet and the (ant)arctic multiplet we
also put the graded spinor derivatives $\na^4_1 , \na^4_2$ of a
propagator $\langle \U (1) \bar{\U} (2) \rangle$ in the internal lines of 
interaction vertices. The action will contain interactions

\be
\int dx \; d^4 \th \oint {d \z \over 2 \pi i \z} {1 \over n! m!}
 \left[ f(\z) \U^n \bar{\U}^m + \overline{f} \({- 1 \over \z}\) 
 \bar{\U}^n \U^m \right]
\ee

\ni
giving vertex factors

\be
\int d^4 \th \oint {d \z \over 2 \pi i \z} \left[ f(\z) \U^q 
 \( \na^4 \)^{n-q} \bar{\U}^{p} \( \na^4 \)^{m-p} + 
 \overline{f} \({-1 \over \z}\) \bar{\U}^{q} 
 \( \na^4 \)^{n-q} \U^p \( \na^4 \)^{m-p} \right] \ .
\ee

\subsection{Diagram construction rules}

Once the explicit form of the propagator and interaction vertices is
known, the rules for diagram construction are easily derived from the
path integral definition of $n$-point functions. As usual, the overall
numerical factors come from the differentiation of the path integral
with respect to $n$ sources, and the different combinatorial
possibilities for connecting the lines of our vertices.

 Once we have constructed a diagram by connecting vertex lines through
propagators, we can extract an overall $\na^4_i$ for each vertex $i$
and use it to complete the restricted superspace measure on that vertex 
to a full $N=2$ superspace measure. This is analogous to completing
the chiral measure of a superpotential interaction to a full $N=1$
superspace measure.

The strategy once we have constructed a given diagram and completed
the $N=2$ superspace measure in all its vertices, is the same as in the 
$N=1$ case \cite{book}: we integrate by parts the $\na$ operators 
acting on some propagator to reduce it to a bare $\d^8 (\th_{ij})$. 
This can be integrated over $\th_i$ or $\th_j$ to bring the vertices
$i$ and $j$ to the same point in $\th$-space. Analogously to the
$N=1$ formalism, we reduce 

\be
 \d^8 (\th_{12}) \na^4_1 \na^4_2 \d^8 (\th_{12})  =
 (\zl -\zZ)^4 \d^8 (\th_{12}) \label{delta_id} \ .
\ee

\ni
Also any number of $\na$'s less than $8$ acting on two
$\d_{12}$ vanishes

\be
 \d^8 (\th_{12}) \na^n_2 \na^4_1 \d^8 (\th_{12}) = 0 , \;\;\; n \neq 4
\ee

\ni
and any number larger than $8$ has to be reduced to $\na^4_{i..j}
\na^4_{m..n}$ by using the anticommutation relations 

\be
 \{ \na_1^{\a}, \bar{\na}_2^{\da} \} = i (\zl - \zZ) \pa^{\a \dot{\a} }. 
\label{n2_ids}
\ee

Transfer rules and integration by parts are used until the
$\th$ dependence has been simplified. The result of these algebraic
manipulations is the integral of a function local in the Grassmann 
coordinates. If we then perform
the contour integrals corresponding to each vertex and integrate on the
Grassmann coordinates of the second supersymmetry, the result must be
the same as that that obtained using the $N=1$ formalism for the
$\z$-coefficient superfields in the hypermultiplet. This is guaranteed
by our construction of the $N=2$ formalism, since the calculations on
that superspace only amount to an integration by parts of the second
supersymmetry spinorial derivatives.

 If we apply this reduction process to the diagrams of the massless
hypermultiplet self-interacting theory, the final amplitude is a 
function of the projective hypermultiplet integrated with the full 
superspace measure. This is the basis of an important 
nonrenormalization theorem: there cannot exist
ultraviolet corrections proportional to the original action $\int d^4 \th
\eta^n$, which is integrated with a restricted superspace 
measure\footnote{After having completed this manuscript we became aware
of related work on \cite{request_ref} }.

 There might conceivably be infrared divergent corrections analogous 
to the $N=1$ infrared divergences contributing to the superpotential

\be
 \int d^4 \th \; \phi^{n-1} {D^2 \over \Box} \phi = \int d^2 \th \; 
 \phi^{n-1} {\bar{D}^2 D^2 \over \Box} \phi = \int d^2 \th \phi^n \ .
\ee

\ni
For example, in the $N=2$ case infrared divergent corrections to the 
$O(4)$ multiplet kinetic action would be of the form

\be
 \int d^8 \th \oint {d \z \over 2 \pi i \z} \eta O \eta = 
 \int d^8 \th \oint {d \z \over 2 \pi i \z} \Psi \na^4 O \eta =
 \int d^8 \th \oint {d \z \over 2 \pi i \z} \Psi {\eta \over \z^2} \ ,
\ee

\ni
where the operator $O$ satisfying this identity is  

\be 
 O = {1 \over \z^2} \oint {d \z_0 \over 2 \pi i \z_0} 
 {\na^4_0 \over (\z - \z_0)^2 \Box^2} \ .
\ee

\ni
It seems unlikely that we can produce such corrections, because the
external $\eta$ fields of $n$-point functions are evaluated at different
positions in $\z$-space

\be 
 \langle \eta_1 \dots \eta_n \rangle = \oint {d \zl \over 2 \pi i \zl} 
 \dots {d \z_n \over 2 \pi i \z_n} \eta (\z_1) \dots \eta (\z_n)
 f(\zl, \dots, \z_n) \ .
\ee

\ni
To obtain a correction to the two point function which is local in
$\z$, we need a $\d$-function in the complex plane as a result of
the loop diagrams

\be
 \oint {d \zZ \over 2 \pi i \z} \;
  \d_{(0)}^{(4)}(\zl, \zZ) \eta (\z_1) \eta (\z_2) = \eta^2 (\zl) \ .
\ee

\ni
From the propagators we obtain delta functions $\d_{(0)}^{(2)} (\zl, \zZ)
$ and from reducing $\d$-functions as in (\ref{delta_id}) we
obtain factors $(\zl -\zZ)^2$. Using anticommutation relations
(\ref{n2_ids}) when there are more than eight $\na$ operators acting on a
$\d$-function we also get factors $(\zl -\zZ)$. With such factors
it is not possible to obtain the proposed delta function, so
it seems that no infrared corrections to the original action are
possible. This arguments are also valid for the finite $O(2p)$
multiplets, but for the arctic multiplets a more careful analysis is
needed.

 An interesting observation is that tadpole diagrams proportional to 
$\eta$ and produced by a point-splitted three point vertex for example (or 
seagull diagrams coming from a four point vertex) 

\be
 \int d^4 \th_1 \oint {d \zl \over 2 \pi i \zl} {\eta (\z_1) \eta (\zl) 
 \over \zl^{3p} } 
 \( \int d^8 \th_2 \; \d^8 (\th_2 - \th_1) \oint {d \zZ \over 2 \pi i \zZ} 
 \d_{(0)}^{(2p)} (\zl, \zZ) \eta (\zZ) \)
\ee

\ni
vanish upon performing the contour integral of the reduced propagator

\beqs
& & \int d^8 \th_2\oint {d \zZ \over 2 \pi i \zZ} \;
 \d_{(0)}^{(2p)} (\zl, \zZ) \; \sum_{n=0}^{2p-2} \zl^n \zZ^{2p-2-n}
 \( \d^8 (\th_1 - \th_2) 
 {\na^4_1 \na^4_2 \over (\zl - \zZ)^2} \d^8 (\th_2 - \th_1) \) \nonumber \\
& & = \oint {d \zZ \over 2 \pi i \zZ} \; \d_{(0)}^{(2p)} (\zl, \zZ) \; 
    \sum_{n=0}^{2p-2} \zl^n \zZ^{2p-2-n} \; (\zl - \zZ)^2 = 0 \ .
\eeqs

\ni
This happens independently of the vanishing of the momentum loop 
integral in dimensional
regularization. Thus the formalism automatically implements the
absence of quadratic divergences in $N=2$ supersymmetry.

 The most striking novelty we find in the diagrams of complex
multiplets corresponds to the additional pole in the convergent limit
(\ref{n2_propupsi12}) of the (ant)arctic multiplet propagator.
Integrating the complex coordinates of two vertices connected by a
complex propagator, involves resolving an ambiguity that arises in the
complex integration. That is because expanding the exponential of the
interaction Lagrangian in the path integral to $n$th order, we find $n$
complex integrations around the same complex contour. Since the
contours are completely overlapping, we do not know if the additional
pole of the (ant)arctic propagator is inside or outside the
contour

\be
\oint_C {d \zl \over (2 \pi i) \zl} \oint_C {d \zZ \over  (2 \pi i) \zZ} 
 \; f( \zl, \zZ) {1 \over \zl - \zZ} \ .
\ee

\ni 
We can give a prescription to integrate on $n$ infinitesimally
separated and concentric contours, so that the  arctic multiplet is
always connected to the antarctic one of the next surrounding
contour. The convergent limit of geometric series in the
arctic-antarctic propagator is then justified for a tree level
diagram. However if we try to construct a 1-loop diagram connecting
the vertices of maximum and minimum contours through their respective
arctic and antarctic multiplets, the propagator will have a divergent
geometric series. A prescription to compute this apparently
ill-defined expression will be presented in a future
publication \cite{n2_effec_action}.

The alternative is to use the generic form of the propagator
(\ref{n2_propupsi12}) where the geometric series has not been replaced
by the convergent limit. We perform the diagram algebra and contour
integrals using an $O(k)$ propagator, and only at the end of
the process we take the limit $k \rightarrow \infty$. It is completely
straightforward and well defined. It gives the same result as that 
obtained with the prescription mentioned before, up to an overall factor of 
two. This is easy to understand because in the complex $O(k)$
propagator we have twice as many physical degrees of freedom as in the 
(ant)arctic multiplet.

\section{Duality between the real $O(2p)$ and (ant)arctic multiplet} 

As we have mentioned, the action for the real $O(2p)$ multiplet and the
action for a real combination of the arctic and antarctic multiplets can
be made dual to each other. For $p=1$ the duality between the tensor
multiplet and the (ant)arctic hypermultiplet exchanges a physical real
$N=1$ linear superfield by a complex $N=1$ chiral field and its
conjugate, and a physical chiral field by a complex linear one.
For $p>1$ the duality relates two off-shell descriptions of the 
hypermutiplet, in which only auxiliary fields are exchanged (see the 
analogous comment for the $N=1$ duality after eq. (\ref{n1_duality}) ). 
Including self-interactions, in the case $p>1$ the duality can be used to 
give different descriptions of the same $\sigma$-model, though there 
may be problems defining the correct contour of integration for the 
(ant)arctic multiplet $\sigma$-model.   

The duality is manifest in the following parent action 

\be
\int dx d^4 \th \oint {d \z \over 2 \pi i \z} \; \;  
 {1 \over 2} (-)^p X^2 
 + (\z^{p-1} \U +(- {1 \over \z})^{p-1} \bar{\U}) X +
 X ({\na^4 \over \z^2} J + {\na^4 \over \z^2} \bar{J} )
\label{pparent_action}
\ee

\ni
where $\U$ and $\na^4 J$ are arctic multiplets, while $X$ is a
tropical multiplet.

In the path integral of the theory, we can integrate out 
$\U$ and $\bar{\U}$. Performing the contour integral
we obtain the following $N=1$ constraints for the coefficient 
superfields:

\beqs 
D_\a X_{-p} = 0 & & D^2 X_{-p+1} = 0	\nonumber	\\
 \bar{D}_{\da} X_p = 0 & & \bar{D}^2 X_{p-1} = 0 \nonumber \\  
& X_n = 0 , \forall |n| > p &
\eeqs

\ni
thus reproducing the real $O(2p)$ multiplet free action for $X
\rightarrow \eta / \z^p$. As before, the source action gives $N=1$ 
unconstrained sources coupled to
the nonzero coefficient superfields of $\eta$.

On the other hand, we can integrate out the real superfield $X$ by 
completing squares on it. Using the real tropical multiplet and (ant)arctic 
multiplet prepotentials to write the action (\ref{pparent_action}) with 
the full $N=2$ superspace measure, the result is

\be
\int dx d^8 \th \oint{d \z \over 2 \pi i \z} {(-)^{p-1}  \over 2}
 \( \z^{p-1} \psi +  { \bar{\psi} (-)^{p-1}  
 \over \z^4 \z^{p-1} }  + { J + \bar{J} \over \z^2} \) 
 \z^2 \na^4 \( \z^{p-1} \psi + {\bar{\psi} (-)^{p-1}  
 \over \z^4 \z^{p-1}} + { J + \bar{J} \over \z^2} \) .
\ee

\ni
The contour integration selects the kinetic action of the (ant)arctic
multiplet

\be 
 \int dx \; d^8\th \oint {d \z \over 2 \pi i \z} \;\; \bar{\psi}  
 {\na^4 \over \z^2} \; \psi = \int dx D^2 \bar{D}^2 
 \oint {d \z \over 2 \pi i \z} \U \bar{\U} \ , 
\ee

\ni
a source term 

\beqs
 \int dx \; d^8 \th \oint {d \z \over 2 \pi i \z} \U \bar{\cal J} (\z) 
 + {\cal J} (\z) \bar{\U}  \,  ; & & {\cal J} = 
 \oint {d \z' \over 2 \pi i \z'} \d_{(0)}^{(+\infty)} (\z,\z') 
 {J (\z') + \bar{J} (\z') \over \z'^{p-1} } , \\
& & \bar{\cal J} = \oint {d \z' \over 2 \pi i \z'} 
 \d_{(-\infty)}^{(0)} (\z,\z') (-\z')^{p-1} \( \bar{J} (\z') + J (\z') \)  
  ,   \nonumber 
\eeqs

\ni
and a term quadratic in sources 

\be
\int dx \; d^8 \th \oint {d \z \over 2 \pi i \z} (-)^{p-1} {1 \over 2}
 (J + \bar{J}) { \na^4 \over \z^2 } (J + \bar{J}) \ . 
\label{key_duality}
\ee

\ni 
To find the path integral of this theory we complete squares on 
the (ant)arctic multiplet prepotential and integrate it out. 
The resulting path integral contains the term quadratic in sources, 
plus our expression (\ref{ok_path_int}) with  
an additional factor $(- \z / \z')^{p-1}$,
and the arctic and antarctic sources replaced by a real tropical source.
Inserting the $N=2$ projective superspace projector and the Riemann
sphere delta distribution in the term 
quadratic in sources, we can also write it as a double complex integral

\beqs
\lefteqn{ ln Z_0 [J+ \bar{J}] = } \\
& & = \int dx d^8 \th 
  \oint {d\z \over 2 \pi i \z} \oint {d\z' \over 2 \pi i \z'} \left[
 \( \bar{J} (\z) + J (\z) \) \z^{p-1} \d_{(0)}^{(+\infty)}(\z,\z')
   {\na^4 (\z) \na^4 (\z') \over \Box \z'^2 (\z -\z' )^2 } 
 { \bar{J} (\z') + J (\z') \over (-\z')^{p-1} } \right. \nonumber \\
& & \left. \qquad \qquad \qquad + {(-)^{p-1} \over 2}  
 \( J (\z)+ \bar{J} (\z) \) { \na^4 (\z) \na^4 (\z') \over \z^2
  (\z-\z')^2 \Box} \( J (\z') + \bar{J} (\z') \) 
 \d_{(-\infty)}^{(+\infty)} (\z,\z') \right].  \nonumber
\eeqs

\ni 
We can use this result to find a relation between the propagator of
the real $O(2p)$ multiplet and that of the (ant)arctic
multiplet. Considering now the full path integral of the dual theory,
we differentiate with respect the real source $J + \bar{J}$

\beqs
 \langle X(1) X(2) \rangle & = & \langle 
 {\d \over \d \( J(1) + \bar{J}(1) \)} 
 { \d \over \d \( J(2) + \bar{J}(2) \)} \rangle  \nonumber \\
& = & \( - {\zl \over \zZ} \)^{p-1} \langle \U (1) 
 \bar{\U} (2) 
 \rangle + \(- {\zZ \over \zl} \)^{p-1} 
 \langle \bar{\U} (1) \U (2) \rangle    \label{dual_prop}   \\
& & + \; (-)^{p-1} \; \d_{(-\infty)}^{(+\infty)} (\zl, \zZ)  
 {1 \over \zl^2}
 {\na^4_1 \na^4_2 \over (\zl-\zZ)^2 \Box} \; \d^8 (\th_1- \th_2) 
 \d (x_1 - x_2) \ . \nonumber
\eeqs

\ni
Substituting the form of the (ant)arctic propagator in
(\ref{dual_prop}) and manipulating the Riemann sphere delta functions 
we obtain indeed $\langle \eta(1) \eta(2) \rangle / \zl^p \zZ^p$

\beqs
 \langle X(1) X(2) \rangle & = & { (-)^{p-1} \over \zl^2} 
 \( - \d_{(-\infty)}^{(1-p)} (\zl,\zZ) - \d_{(p+1)}^{(+\infty)} 
 (\zl,\zZ) +  \d_{(-\infty)}^{(+\infty)} (\zl,\zZ) \) \nonumber \\
& & \qquad \qquad \qquad \qquad   \times 
 {\na^4_1 \na^4_2 \over (\zl-\zZ)^2 \Box} \d^8 (\th_1- \th_2) 
 \d (x_1 - x_2)   \\
& = & { (-)^{p-1} \over \zl^2} \; \d_{(2-p)}^{(p)} (\zl,\zZ) 
 {\na^4_1 \na^4_2 \over (\zl-\zZ)^2 \Box} \d^8 (\th_1- \th_2) 
 \d (x_1 - x_2) \ , \nonumber 
\eeqs

\ni
where equating the first line to the last one is understood to apply
when we integrate on the Riemann sphere with the real measure 
$\oint{ d\z \over 2\pi i \z}$. Thus duality gives the correct relation
between the propagators.

\section{Acknowledgments}
UL acknowledges support from NFR grant No 4038-312 and from NorFA, 
grant No 96.55.030-O. FGR, MR and SW acknowledge partial support from
NSF grant No Phy 9722101. MR thanks the ITP at Stockholm and UL for
their hospitality. FGR thanks the ITP at Stockholm and RvU for their
hospitality.

\end{document}